\documentclass[12pt]{article}

\usepackage{latexsym}
\usepackage{amsfonts}
\usepackage{amssymb}
\usepackage[all]{xy}

\begin{document}
\newcommand{\proof}{{\em Proof. }}
\newcommand{\ed}{\hfill Q.E.D.\\}
\newcommand{\we}{\wedge}
\newcommand{\prt}{\partial}
\newcommand{\sd}[1]{\Sigma_{#1}}
\newcommand{\sti}{\Sigma_{\tau_i}}
\newcommand{\ebd}[1]{{\bf e}_{(#1)}}
\newcommand{\egd}[2]{e^{#1}_{(#2)}}
\newcommand{\ad}[1]{a_{#1}}
\newcommand{\tmn}{T^{\mu\nu}}
\newcommand{\tgg}[2]{T^{{#1}{#2}}}
\newcommand{\tgn}[2]{T^{(#1)(#2)}}
\newcommand{\ebg}[1]{{\bf e}^{(#1)}}
\newcommand{\mt}[3]{{\cal M}^{{#1}{#2}{#3}}}
\newcommand{\md}[2]{M^{{#1}{#2}}}
\newcommand{\mmnl}{\mt\mu\nu\lam}
\newcommand{\mnl}{\md\nu\lam}
\newcommand{\mbnl}{M^{B\nu\lam}}
\newcommand{\mcnl}{M^{C\nu\lam}}
\newcommand{\dbk}{(D\times B)^k}
\newcommand{\dbp}{\frac12(D^2+B^2)}
\newcommand{\rtf}{R(\theta,\phi)}
\newcommand{\ritf}{R_{(i)}(\theta,\phi)}
\newcommand{\lar}{\log(a\rtf)}
\newcommand{\lari}{\log(a\ritf)}
\newcommand{\oi}{{\cal O}_i}
\newcommand{\uni}{u^{\nu}_{(i)}}
\newcommand{\pbn}{p^{B\nu}}
\newcommand{\pcn}{p^{C\nu}}
\newcommand{\fbi}{f^B_{(i)}}
\newcommand{\fci}{f^C_{(i)}}
\newcommand{\dss}{\frac12[(D^s)^2]_s}
\newcommand{\oo}{{\cal O}}
\newcommand{\dt}[1]{\frac{d #1}{d\tau}}
\newcommand{\st}{\sd\tau}
\newcommand{\pti}{P_{\tau_i}}
\newcommand{\pni}{P_{(i)}}
\newcommand{\ptz}{P_{\tau}}
\newcommand{\ba}{\backslash}
\newcommand{\dsm}{d\sigma_{\mu}}
\newcommand{\fb}{f^B}
\newcommand{\fc}{f^C}
\newcommand{\et}{\label}
\newcommand{\rf}[1]{(\ref{#1})}
\newcommand{\re}{\ref}
\newcommand{\ab}{\mbox{{\bf a}}{}}
\newcommand{\bb}{\mbox{{\bf b}}{}}
\newcommand{\ub}{\mbox{{\bf u}}{}}
\newcommand{\eb}{\mbox{{\bf e}}{}}
\newcommand{\te}{\tilde}
\newcommand{\al}{\alpha}
\newcommand{\ga}{\gamma}
\newcommand{\eps}{\epsilon}
\newcommand{\es}{s}
\newcommand{\lam}{\lambda}
\newcommand{\si}{\sigma}
\newcommand{\de}{\delta}
\newcommand{\dd}{\Delta}
\newcommand{\la}{\lambda}
\newcommand{\Si}{\Sigma}
\newcommand{\ds}{d\Sigma}
\newcommand{\be}{\begin{equation}}
\newcommand{\ee}{\end{equation}}
\newcommand{\ra}{\rightarrow}
\newcommand{\is}{\int_{S^2}}
\newcommand{\su}{\subset}
\newcommand{\pua}{P^{\nu}_{(i)}(\ub_{(i)},\ab_{(i)})}
\newcommand{\mua}{M^{\nu\lam}(\ub,\ab)}
\newcommand{\ui}{{\cal U}_i}
\newcommand{\uc}{{\cal U}}
\newcommand{\pu}{P^{\nu}(\ub,\ab)}

\title{\bf {\huge Born renormalization\\ in classical Maxwell
electrodynamics}}
\author{Jerzy Kijowski${}^1$ and Piotr Podle\'s${}^2$\\
${}^1$ Center for Theoretical Physics, Polish Academy of
Sciences,\\ Aleja Lotnik\'ow 32/46, 02--668 Warszawa, Poland\\
${}^2$ Department of Mathematical Methods in Physics,\\ Faculty of
Physics,  Warsaw University,\\ Ho\.za 74, 00--682 Warszawa,
Poland}
\date{}
\maketitle
\begin{abstract}
We define and compute the renormalized four--momentum of the
composed physical system: classical Maxwell field interacting with
charged point particles. As a `reference' configuration for the
field surrounding the particle, we take the Born solution. Unlike
in the previous approach \cite{EMP} and \cite{GKZ}, based on the
Coulomb `reference', a dependence of the four--momentum of the
particle (`dressed' with the Born solution) upon its acceleration
arises in a natural way. This will change the resulting equations
of motion. Similarly, we treat the angular momentum tensor of the
system.
\end{abstract}

\section{Introduction}

Classical, relativistic electrodynamics is unable to describe
interaction between charged particles, intermediated by
electromagnetic field. Indeed, typical well posed problems of the
theory are of the contradictory nature: either we may solve
partial differential equations for the field, with particle
trajectories providing sources (given {\em a priori} !), or we may
solve ordinary differential equations for the trajectories of test
particles, with fields providing forces (given {\em a priori} !).
Combining these two procedures into a single theory leads to a
contradiction: Lorentz force due to self-interaction is infinite
in case of a point particle. Replacing point particle by an
extended object is not a good remedy for this disease because it
requires a field-theoretical description of the interior of the
particle (rigid spheres do not exist in relativity!). This means
that the three degrees of freedom of the particle must be replaced
by an infinite number of degrees of freedom of the matter fields
constituting the particle. Moreover, a highly nonlinear model for
the interaction of these fields must be chosen in order to assure
the stability of such an object. As a consequence, there is no
hope for an effective theory.

There were many attempts to overcome these difficulties. One of
them consists in using the Lorentz--Dirac equation, see
\cite{Dirac},\cite{Haag},\cite{Rohr}. Here, an effective force by
which the retarded solution computed for a given particle
trajectory acts on that particle is postulated (the remaining
field is finite and acts by the usual Lorentz force).
Unfortunately, this equation has many drawbacks (cf. Section 8 of
\cite{EMP}). See also \cite{WF} for another approach to this
problem.

In papers \cite{EMP} and \cite{GKZ} a mathematically consistent
theory of the physical system ``particle(s) + fields'' was
proposed, which overcomes most of the above difficulties even if
some problems still remain. The theory may be defined as follows.
We consider a system consisting of charged point particles and the
electromagnetic field $f_{\mu\nu}$. We always assume that the
latter fulfills Maxwell equations with Dirac ,,delta-like''
currents defined uniquely by the particle trajectories. Given such
a system, we are able to define its total ,,renormalized
four-momentum''. For a generic choice of fields and particle
trajectories this quantity {\em is not conserved}. Its
conservation is an additional condition which we impose on the
system. It provides us the missing ,,equations of motion'' for the
trajectories and makes the system mathematically closed (cf.
\cite{GKZ}).

Definition of the renormalized four-momentum of the system
composed of fields and particles, proposed in \cite{EMP}, was
based on the following reasoning. Outside of the particles, the
contribution to the total four--momentum carried by the Maxwell
field $f_{\mu\nu}$ is given by integrals of the Maxwell
energy--momentum tensor--density
\be
T^{\mu\nu}=T^{\mu\nu}(f)=\sqrt{-g}(f^{\mu\lam}f^{\nu}_{\lam}-\frac14
g^{\mu\nu}f^{\kappa\lam}f_{\kappa\lam}) \et{1a} \ee over a
space--like hypersurface $\Sigma$ (the notation is prepared for
working in curvilinear coordinates). Unfortunately, the total
integral of this quantity is divergent because of the field
singularities at the particle's positions. The idea proposed in
\cite{EMP} is to consider for each particle a fictitious
``reference particle'' which moves uniformly along a straight line
tangent to the trajectory of the real particle at the point of
intersection with $\Sigma$. The constant velocity $\ub$ of this
hypothetical particle is thus equal to the instantaneous velocity
of the real particle at the point of intersection. Give a label
$(i)$ to each of those hypothetical particles and consider the
corresponding Coulomb field $f^C_{(i)}$ boosted to velocity
$\ub_{(i)}$. In the rest frame of the $(i)$-th particle, the
magnetic and electric components of this field may be written as
\begin{equation}\label{1b}
  B^C_{(i)}=0,\qquad
  (D^C_{(i)})^k=\frac{e_{(i)}}{4\pi}\frac{x^k}{r^3}\  .
\end{equation}
The ``reference particle'' has the same charge $e_{(i)}$ and the
same rest mass $m_{(i)}$ as the real particle. By the mass we
mean, however, not the  ``bare mass'', which must later be
``dressed'' with the energy of its Coulomb tail (which always
leads to infinities during renormalization procedure), but the
total energy of the composed system ``particle + field'' at rest.
Hence, the total four--momentum of the $i$'th reference particle
(together with its field $f^C_{(i)}$) equals
\begin{equation}\label{tot-ren}
  p_{(i)}^{\nu} = m_{(i)}u^{\nu}_{(i)} \ .
\end{equation}

Now, to define the renormalized four-momentum $p^{C\nu}$ carried
by the particles and the field $f_{\mu\nu}$ surrounding them, we
split the energy-momentum density $T(f)$ into the sum of the
reference densities $T(f^C_{(i)})$ and the remaining term.
According to \cite{EMP}, the remaining term is integrable (more
strictly, the principal value of the integral exists), while
$T(f^C_{(i)})$ terms are already ``taken into account'' in the
four--momenta $m_{(i)}\ub_{(i)}$ of the particles (computed at the
points of intersection). Hence, the ``Coulomb--renormalized
four--momentum'' of the system is defined by the following
formula:
\begin{equation}\label{2}
  p^{C\nu}:=P\int_{\Si}\left[ T^{\mu\nu}(f)-\sum_i
  T^{\mu\nu}(f^C_{(i)})\right]d\si_{\mu} + \sum_i
  m_{(i)}u^{\nu}_{(i)} \ .
\end{equation}
It was proved in \cite{EMP} that $p^{C\nu}$ depends on $\Si$ only
through the points $A_i$ of intersection of $\Si$ with the
trajectories. Next, one postulates that $p^{C\nu}$ doesn't depend
on those points. This condition implies the dynamics of the
particles \cite{EMP} and makes the evolution of the system unique
(cf. \cite{GKZ}).

The above theory is not completely satisfactory, because the
subtraction of $T(f^C_{(i)})$ in \rf2 kills only terms which
behave like $r^{-4}$,  while the $r^{-3}$-terms remain in \rf2 and
are integrated with $r^2dr$ (for simplicity we assume here that
$\Si$ near the particle corresponds to $x^0=$ const. in the rest
frame). This phenomenon is implied by the analysis of the Maxwell field
behaviour in the vicinity of the particle, cf. \rf1 or Section 5
of \cite{EMP}. It leads to logarithmic divergencies which
disappear only due to the principal value sign $P$ in front of the
integral (\ref{2}). That sign means that we first compute the
integral over $\Sigma \backslash {\cal U}$, where ${\cal
U}=\cup\uc_i$ and $\uc_i$ is a small {\em symmetric} neighbourhood
of the i-th particle and then we pass to the limit with $\uc_i$
shrinking to the point: $\uc_i \rightarrow A_i $. The symmetry is
necessary to kill the $r^{-3}$-term under integration because it
is anti-symmetric.

The main result of the present paper is a new, improved
renormalization procedure, which does not rely on the symmetry of
$\uc_i$.  We call this new procedure a {\em Born renormalization},
because the Coulomb reference for a moving particle, matching only
its velocity, is here replaced by the Born solution, matching both
the velocity and the acceleration of the particle.

We are going to prove in the sequel, that the four-momentum
defined {\em via} the Coulomb--renormalization is a special case
of the result obtained {\em via}  Born--renor\-ma\-li\-za\-tion,
while the physical interpretation of the latter is more natural:
all the integrals occurring here are uniquely defined without any use
of the principal value sign. Moreover, the ultra-local dependence
of the four-momentum upon the acceleration of the particle,
implied by the Born renormalization, will change the equations of
motion of the particles. That dependence may be also a key to the
instability problem of the theory (with an appropriate dependence
of the involved functions on the acceleration).  We prove in
Section \ref{relation} that, disregarding this dependence, we
recover the previous Coulomb-renormalized formulae.

Our results are based on an analysis of the behaviour of the
Maxwell field in the vicinity of the particles done in papers
\cite{JKMK}, \cite{LW} (cf. also \cite{Dirac}). Although the
asymptotic behaviour of the radiation field far away from the
sources may be found in any textbook, the ``near-field'' behaviour
is less known. The main observation is that -- for any choice of
particle trajectories -- the difference between the retarded and
the advanced solution is bounded (in the vicinity of the
particles). Hence, we restrict our considerations to the fields
which differ from the particle's retarded (or advanced) field by a
term which is bounded in the vicinity of that particle. We also
assume that the field at spatial infinity, i.e. for $r\ra\infty$, 
is at most of the order of $r^{-2}$. Fields fulfilling those
requirements are called {\em regular}. In
the particle's rest frame, regular fields have the following
behaviour near the particle (cf. (25) of \cite{EMP}):
\begin{equation}\label{1}
  B^k=\tilde B^k,\quad D^k=(D^s)^k+\te D^k,\quad
  (D^s)^k=\frac{e}{4\pi} \frac{x^k}{r^3} - \frac{e}{8\pi r} \left(
  a_{i}\frac{x^ix^k}{r^2}+a^{k}\right) \ ,
\end{equation}
where $\te B,\te D$ are bounded and $a^k$ are the components of
the acceleration of the particle. Above formulae may be proved for
the retarded field using Lienard--Wiechert potentials (cf.
\cite{Dirac},\cite{LW},\cite{JKMK})\footnote{We use this
opportunity to correct a missprint in formulae (77)--(79) of
\cite{LW}: the right hand sides should be multiplied by $r$ and
the indices below $B$ on the left hand sides should be increased
by one. Correct formulae for the arithmetic mean of the retarded
and the advanced fields may be found in \cite{JKMK}. }. Hence,
they are valid for all regular fields.

Our paper is organized as follows. In Sections 2 and 3 we recall
and investigate the Fermi--propagated system of coordinates and
the Born solution. In Sections 4--7 we restrict ourselves (for
simplicity) to the case of a single particle interacting with the
field (A straightforward generalization of these results to the
case of many particles is given in Section 8. This generalization
does not require any new ingredient because interaction between
particles is intermediated {\em via} linear Maxwell field.) In
Section 4 we define the Born--renormalized four--momentum
$p^{B\nu}$ of the system and prove that it depends on the
hypersurface $\Si$ through the point of intersection with the
trajectory only. In Section 5 we assume that $\Si$ near the
trajectory coincides with the $x^0=$ const. in the Fermi system
which allows us to find an explicit expression for $p^{B\nu}$. In
Section 6 we compare $p^{B\nu}$ with the Coulomb--renormalized
$p^{C\nu}$ of \cite{EMP}. The difference of the two is a function
of four--velocity and acceleration at the point of intersection.
In Section 7 we extend the results of Sections 4--6 to the case of
the angular momentum tensor. The fall-off conditions at spatial
infinity and technical details of the proofs are presented in
Appendices.

We stress that our approach to renormalization never uses any
cancellation procedure of the type "$+\infty - \infty$". Here,
everything is finite from the very beginning and the point
particle is understood as a mathematical model, approximating a
realistic, physical particle which is assumed to be extended. To
formulate such a model one has to abandon the idea of a point
particle ``floating over the field'' but rather treat it as a tiny
``strong field region'' (its internal dynamics is unknown but --
probably -- highly nonlinear), surrounded by the ``week field
region'', governed by the linear Maxwell theory. The strong field
region (particle's interior) interacts with the field {\em via}
its boundary conditions. In other words: the idea to divide
``horizontally'' the total energy of the system into: 1) the
``true material energy'' + 2) the free field energy and, finally,
3) the interaction energy which adds to the previous two
contributions, must be rejected from the very beginning. Such a
splitting, which is possible for linear systems, makes no sense in
case of a realistic particle. In our approach, only ``vertical''
splitting of the energy into contributions contained in disjoint
space regions, separated by a chosen boundary, makes sense because
of the locality properties of the theory.

The main advantage of the theory constructed this way is its
universality: the final result does not depend upon a specific
structure of the particle's interior, which we want to
approximate. Moreover (what is even more important!), it does not
depend upon a choice of the hypothetical ,,boundary'' which we
have used to separate the the strong field region from the weak
field region: the only assumption is that it is small with respect
to characteristic length of the external field.

\section{The Fermi--propagated system}

In this Section we recall and investigate the properties of the
Fermi--propa\-ga\-ted system of coordinates. It is a non--inertial
system such that the particle is at rest at each instant of time.
The use of the Fermi system simplifies considerably description of
the field boundary conditions in the vicinity of the particle,
given by \rf1 and \rf{1b}. The price we pay for this
simplification is a bit more complicated (with respect to the
inertial system) description of the field dynamics, cf.
\cite{EMP}, \cite{JKMK}.

Let $y^{\lam}$, $\lam=0,1,2,3$, denote the (Minkowski) spacetime
coordinates in a fixed inertial (`laboratory') system. By ${\bf
f}_{\lam}=\frac{\partial}{\partial y^\lam}$ we denote the
corresponding orthonormal basis for the metric tensor $\eta={\rm
diag}(-,+,+,+)$. Let $q^{\lam}(t)=(t,q^k(t))$ be a particle's
trajectory and $\tau=\tau(t)$ be the particle's proper time. Then
$\frac{d\tau}{dt}=(1-v^2)^{1/2}$ where $v^k=\dot q^k$ (dot denotes
the derivative w.r.t. $t$). The normalized four--velocity is given
by: $\ub=\dt q=((1-v^2)^{-1/2},(1-v^2)^{-1/2}v^k)$ and the
particle's acceleration $\ab=\dt\ub=\frac{d^2q}{d\tau^2}$.
Clearly, $(\ub|\ab)=0.$

We define the rest-frame space $\sd\tau$ as the hyperplane
orthogonal to the trajectory (i.e. to $\ebd0=\ub$) at $q(t)$.
Choose any orthonormal basis $\ebd l$, $l=1,2,3$, in $\sd\tau$,
such that $\ebd\mu$ are positively oriented.  Thus
$(\ebd\al|\ebd\beta)=\eta_{\alpha\beta}$. Denote by $\ebd
l(t)=(c_l(t),d^k_l(t))$, $l=1,2,3$ the laboratory components of
the triad. We define a new system of coordinates
$x^{\mu}=(\tau,x^l)$ putting $ y^{\lam}=q^{\lam}(t)+x^l\egd\lam
l(t) $. This is only a local system, defined in a vicinity of the
trajectory. For fixed $\tau$ (or $t$), $y$ cover the entire
$\sd\tau$ and the particle remains always at the origin $x^l=0$.
In coordinates $(x^\mu)$ the metric tensor equals
$g_{\mu\nu}=(\frac\prt{\prt x^{\mu}}|\frac\prt{\prt x^{\nu}})$
where $\frac\prt{\prt x^{\mu}}\equiv\frac{\prt y}{\prt x^{\mu}}$.
In particular, $\frac\prt{\prt\tau}=\ub+x^l\dt{\ebd l}$,
$\frac\prt{\prt x^l}=\ebd l$. Thus $g_{kl}=\de_{kl}$.
Orthogonality condition $({\ebd l}|\ub)\equiv 0$ implies the
following identity: $\dt{}(\ebd l|\ub)=0$ which means that
$(\dt{\ebd l}|\ub)=-(\ebd l|\dt\ub)=-(\ebd l|\ab)=-a_l$, where
$a_l\ebd l=\ab$.

{}Fermi frame is defined by the following constraint imposed on
the triad $\ebd l$: $g_{0l}\equiv N_l=0$. This implies that
$\dt{\ebd l}$ is proportional to $\ub$, $\dt{\ebd l}=a_l\ub$ and
determines the propagation of $\ebd l$ uniquely (provided they are
given for $t=t_0$) and consistently (one has
$\dt{}(\ebd\mu|\ebd\nu)=0$). This condition implies $\dot
c_l=a_l$, $\dot d^k_l=v^ka_l$. Moreover, one has
$\frac\prt{\prt\tau}=N\ebd0$, where $N=1+a_lx^l$. Thus
$g_{00}=(N\ebd 0|N\ebd 0)=-N^2$ (i.e., N is the lapse function).

In this Fermi--propagated system the field $f$ is related to the
electric and magnetic fields by (cf. (5)--(6) of \cite{JKMK})
\[ f^{0k}=N^{-1}D^k,\qquad f^{kl}=\eps^{klm}B_m. \]
Sometimes it is more convenient to use nonholonomic field
coordinates $f^{(\al)(\beta)}$, calculated w.r.t. the tetrad
$\ebd\alpha$. They are related to $f^{\mu\nu}$ by
$f^{\mu\nu}=\egd\mu\al\egd\nu\beta f^{(\al)(\beta)}$ where
$\mu,\nu$ are taken w.r.t. $(y^{\lam})$ or, alternatively, w.r.t.
$(x^{\lam})$. In the latter case one has $\egd k l=\de^k_l$, $\egd
k0=\egd 0k=0$, $\egd00=N^{-1}$, which gives $f^{(0)(k)}=D^k$,
$f^{(k)(l)}=\eps^{klm}B_m$, like in the laboratory frame. Also
$g_{(\al)(\beta)}=(\ebd\al|\ebd\beta)=\eta_{\al\beta}$. Thus
$\tgn\al\beta$ has the same form as in the laboratory: \be
\left.\begin{array}{l} \tgn00=\dbp,\quad \tgn0k=\tgn k0=\dbk,\\
\tgn k l=-D^kD^l-B^kB^l+\frac12\de^{kl}(D^2+B^2).
\end{array}\right\} \et{3}\ee
We shall use the following

 {\em Proposition.} When integrating over $\oo\subset\sd\tau$,
one can put (in any system of coordinates)
\[ \egd\mu\al\dsm=\de_{\al0}d\Si \]
where $d\Si$ is the volume element for $\st$ and $\dsm$ are the
basic three--volume forms.

{\em Proof.} Taking the laboratory frame,
$\egd\mu\al\dsm=\egd\mu\al\frac\prt{\prt y^{\mu}}\rfloor dy^0\we
dy^1\we dy^2\we dy^3=\ebd\al\rfloor\ebg0\we\ebg1\we\ebg2\we\ebg3$
equals $\ebg1\we\ebg2\we\ebg3=d\Si$ for $\al=0$, but for
$\al\neq0$ it contains $\ebg0$, hence it vanishes when we
integrate over $\oo\subset\st$.\ed

Now consider the laboratory frame. On each hypersurface $\st$ we
introduce coordinates $\te
y^{\lam}=y^{\lam}-q^{\lam}(t)=x^l\egd\lam l$ calculated w.r.t. the
particle and we decompose the angular momentum tensor--density \be
\mmnl= y^{\nu}\tgg\mu\lam-y^{\lam}\tmn \et{4}\ee as follows: \be
\mmnl=\te\mmnl+q^{\nu}\tgg\mu\lam-q^{\lam}\tmn, \et{6aa}\ee \be
\te\mmnl= \te y^{\nu}\tgg\mu\lam-\te y^{\lam}\tmn. \et{6ab}\ee
Here $\tilde{\cal M}$ computed at $y$ is the angular momentum
tensor--density w.r.t. the position of the particle $q(t)$ such
that $y$ belongs to the hyparplane $\st$ with $q(t)$ at its
origin.

Integrating over $\oo\subset\st$ one may use the nonholonomic
coordinates $T^{(\al)(\beta)}$ (cf. Proposition and \rf3):
\[
\tmn\dsm=\egd\mu\al\egd\nu\beta\tgn\al\beta\dsm=\egd\nu\beta\tgn0\beta
d\Si, \] \be \tmn\dsm=\egd\nu0\dbp d\Si+\egd\nu k\dbk d\Si,
\et{4a}\ee
\[\te \mmnl\dsm=(x^l\egd\nu
l\egd\mu\al\egd\lam\beta\tgn\al\beta-x^l\egd\lam
l\egd\mu\al\egd\nu\beta\tgn\al\beta)\dsm \]
\[ =x^l(\egd\nu l\egd\lam\beta-\egd\lam l\egd\nu\beta)\tgn0\beta
d\Si, \] \be\left.\begin{array}{l} \te\mmnl\dsm=(\egd\nu
l\egd\lam0-\egd\lam l\egd\nu0)x^l\dbp d\Si+\\ (\egd\nu l\egd\lam
k-\egd\lam l\egd\nu k)x^l\dbk d\Si.\end{array} \right\} \et{4b}\ee

\section{The Born solution}

Consider a uniformly accelerated particle, i.e. $a_l=\rm{const.}$
We have: $\frac{d{\ub}}{d{\tau}}=a_l\ebd l$, $\frac{d{ {\bf
e}}_{(l)}}{d{\tau}}=a_l{\ub}$, which determines its trajectory (a
hyperbola) and its Fermi-propagated system uniquely, provided
$q(0)=0$ and the initial data $a_l$, ${\bf e}_{(l)}$, ${\bf u}$
are given. The propagation may be obtained by action of the
one-parameter group ${\cal G}$ of proper Lorentz transformations
(boosts) on initial data. The group ${\cal G}$ leaves invariant
the point $-e_{(l)}(0)a^l/a^2$, where $a=|\ab|$.

We may use a time-independent 3-rotation from $\ebd l$ to a new
triad ${\bf b}_{(l)}$, such that the acceleration ${\bf a}$ is
proportional to the third axis: $a^l{ {\bf e}}_{(l)}
\equiv{\ab}=a{\bf b}_{(3)}$. Denote the corresponding
Fermi--propagated coordinates by $x^l$ (for ${\bf e}_{(l)}$) and
by $z^l$ (for ${\bf b}_{(l)}$). The spherical coordinates related
to $z^l$ are called $r,\theta,\phi$. The Born solution of Maxwell
equations with a delta-like source carried by the particle (cf.
\cite{Rohr},\cite{Th}, Section 3.3 of \cite{MK} and \cite{Tu})
reduces in these coordinates to the following time--independent
expression: \be D_r = \frac{e}{\pi r^2}
\frac{2+ar\cos\theta}{(a^2r^2+4+4ar\cos\theta)^{3/2}}, \et{5a} \ee
\be D_{\theta}= \frac{e}{\pi r^2}
\frac{ar\sin\theta}{(a^2r^2+4+4ar\cos\theta)^{3/2}}, \et{5b} \ee
\be D_{\phi}=B_r=B_{\theta}=B_{\phi}=0. \et{5c} \ee
 The electric field $D$ is singular not only at $r=0$, where it
behaves as in \rf1 with $\te D$ bounded (cf. \cite{MK}), but also
for $r=2/a$, $\theta=\pi$. It turns out that the solution
describes {\em two} symmetric particles with opposite charges and
opposite accelerations. Actually, the Born solution may be defined
as a unique solution of the problem which is invariant with
respect to the symmmetry group ${\cal G}$ of the problem and
satisfies other natural assumptions (cf. \cite{Tu},\cite{MK}).

The Fermi propagation consists in acting with the Lorentz
rotations (boosts) $g \in {\cal G}$ on the hyperplanes $\st$. This
action leaves the 2-plane $\mathfrak{p}:= \left\{ N = 0 \right\} =
\left\{ z^3 = -\frac1a \right\}$ invariant. The plane splits each
$\st$ into two half--hyperplanes. Denote by
$P_{\tau}=\{x\in\st:z^3>-\frac1a\}$ the one which contains our
original particle situated at $r=0$.

Assume that $\oo\subset P_{\tau}$ is a small region around the
particle described by $r<\rtf$ where the latter is a given
function. In Section 5 we shall need

{\em Proposition.} \be\left.\begin{array}{l} \int_{\ptz\ba\oo}
\dbp d\Si = \int_E \frac{e^2}{2\pi^2 r^2} \frac{\sin\theta dr
d\theta d\phi }{(a^2r^2+4+4ar\cos\theta)^2}\\ =
\frac{e^2}{2\pi^2}\is \{\frac1{16\rtf}+\frac18
a\cos\theta\lar+O(R)\} \sin\theta d\theta d\phi,
\end{array}\right\} \et{9} \ee
where $E=\ptz\ba\oo= \{(r,\theta,\phi):\ \  r\geq\rtf, \ \
r\cos\theta>-1/a\}$ (in spherical coordinates).

\be\left.\begin{array}{l} \int_{\ptz\ba\oo} z^k \dbp d\Si\\ =
\frac{e^2}{32\pi^2}\is \{-\lar \frac{z^k}r +O(R)\} \sin\theta
d\theta d\phi -\frac{e^2}{8\pi}\de_{3k}.\end{array}\right\}
\et{10} \ee

{\em Proof.} By lengthy but standard computations.\ed

The above result can be reformulated using the following
construction. For $G=\sum_k r^k f_k(\theta,\phi)$ we define its
singular part $G_s=\sum_k r^k f_k(\theta,\phi) 1_{[0,d_k]}(r)$
where $d_k=0$ for $k>-3$, $d_k=1/a$ for $k=-3$, $d_k=+\infty$ for
$k<-3$ ($1_B$ is a characteristic function of a set $B$, i.e.
$1_B(y)=0$ for $y\not\in B$, $1_B(y)=1$ for $y\in B$). Then
(cf.\rf1) \be \dss=\frac{e^2}{32\pi^2}
\left(\frac1{r^4}-\frac{2a\cos\theta}{r^3}1_{[0,1/a]}\right),
\et{10p} \ee \be \int_{\st\ba\oo} \dss
\ds=\frac{e^2}{2\pi^2}\is\left\{\frac1{16\rtf}+ \frac18
a\cos\theta\lar\right\} \sin\theta d\theta d\phi, \et{10a}\ee \be
\left[\frac12
z^k(D^s)^2\right]_s=\frac{e^2z^k}{32\pi^2r^4}1_{[0,1/a]},
\et{10aa}\ee
\be
\int_{\st\ba\oo} \left[\frac12 z^k(D^s)^2\right]_s \ds=
-\frac{e^2}{32\pi^2}\is \lar \frac{z^k}r \sin\theta d\theta d\phi.
\et{10b}\ee

\section{The Born-renormalized four-momentum}
\label{Born-renormalization}

Throughout the paper we assume that the particle has no internal
degrees of freedom, i.~e.~it is completely characterized by its
charge $e$ and mass $m$. Consider a regular Maxwell field $f$
consistent with the trajectory of the particle (cf. Section 1). We
fix a point $A$ on its trajectory, corresponding to given values
of the proper time $\tau$, four--velocity $\ub$ and acceleration
$\ab$.

Formula (\ref{2}) for the Coulomb-renormalized four-momentum was
based on the following heuristic picture: A real, physical
particle is an extended object, an exact solution of the complete
system: ``matter fields + electromagnetic field''. The reference
particle (passing through $A$ and moving with the constant
four--velocity ${\bf u}$) is also an exact, stable solution of the
same system, which, moreover, is static (,,soliton-like'').
Outside of a certain small radius $r_0$ the matter fields vanish
and the electromagnetic field reduces to the Coulomb field $f^C$.
Hence, for ${\uc}$ which is very small from the macroscopic point
of view but still big from the microscopic point of view (i.e.
much bigger than the ball $K(A,r_0)$ around the particle), the
total amount of the four-momentum carried by the soliton solution
and contained in ${\uc}$ equals:
\begin{equation}\label{inO}
   p^{C\nu} (\uc)=
  m u^{\nu} - \int_{\st \ba\uc}
  T^{\mu\nu}(f^C)d\si_{\mu} \ .
\end{equation}
The stability assumption means, that for the real particle
surrounded by the field $f$, the amount of the four-momentum
contained in $\uc$ does not differ considerably from the above
quantity, provided $\uc$ is very small with respect to the
characteristic length of $f$. Together with the amount of the
four-momentum contained outside of $\uc$:
\begin{equation}\label{outO}
   p^{\nu} (\st \ba\uc)=
  \int_{\st \ba\uc}
  T^{\mu\nu}(f)d\si_{\mu} \ ,
\end{equation}
quantity (\ref{inO}) provides, therefore, a good approximation of
the total four-momentum of the ``extended particle +
electromagnetic field'' system:
\begin{equation}\label{tot}
   p^{\nu} \simeq
  \int_{\st \ba\uc}\left[
  T^{\mu\nu}(f)d\si_{\mu} -  T^{\mu\nu}(f^C)
  \right] + m u^{\nu} \ .
\end{equation}
Treating the point particle as an idealization of the extended
particle model and applying the above idea, we may shrink $\uc$ to
a point, i.~e.~$\uc \rightarrow A$, with respect to the
macroscopic scale (but keeping $\uc$ always very big with respect
to the microscopic scale $r_0$). This procedure -- in case of many
particles -- gives us precisely formula (\ref{2}).

Now, we assume that also the Born solution has its
``extended-particle version''. More precisely, we assume that the
total system: ``matter fields + electromagnetic field'', admits a
stable, stationary (with respect to the one-parameter group ${\cal
G}$ of boosts) solution, which coincides with the Born field $f^B$
outside of a certain small radius $r_0$ around the particles. This
solution represents a pair of uniformly accelerated particles.
Denote by $\pu$ the amount of the total four-momentum carried by
this solution in the half-hyperplane $\ptz$. Hence, the amount of
the four-momentum contained in $\uc$ equals:
\begin{equation}\label{B-inO}
   p^{B\nu} (\uc)=
  \pu - \int_{\ptz \ba \uc}
  T^{\mu\nu}(f^B)d\si_{\mu} \ .
\end{equation}
Replacing (\ref{inO}) by (\ref{B-inO}) in formula (\ref{tot}), we
obtain the following approximation for the total four-momentum:
\begin{eqnarray}\label{12}
  p^{\nu} & \simeq & \int_{\st \ba\uc}
  \tmn(f)\dsm
  -\int_{\ptz \ba\uc}\tmn(\fb)\dsm  +\pu  \\
  & = &  \int_{\st \ba {\cal O}} \tmn(f)\dsm
  -\int_{\ptz \ba {\cal O}}\tmn(\fb)\dsm   \nonumber\\
   & + & \int_{\cal O\ba\uc} \left[ \tmn(f) - \tmn(\fb) \right] \dsm
  \ + \ \pu, \label{Born-Fermi-U}
\end{eqnarray}
where ${\cal O}$ is a fixed macroscopic neighbourhood of the
particle, contained in $\ptz$ and containing $\uc$. Again,
treating point particle as an idealization of the extended
particle model and applying the above idea, we may pass to the
limit $\uc \rightarrow A$ with respect to the macroscopic scale
(but keeping $\uc$ always very big with respect to the microscopic
scale $r_0$). Unlike in the Coulomb renormalization, the limit
exists without any symmetry assumption about $\uc$, because
$T(f)-T(\fb)$ behaves like $r^{-2}$ in the vicinity of the
particle (due to formulae \rf{1}, \rf{3} and Section 3). Hence, we
obtain the following

{\em Definition.} The renormalized four-momentum of the ``point
particle + electromagnetic field'' system is given (in the
laboratory system) by

\begin{equation}\left.
\begin{array}{rcl}
  \pbn & := & \int_{\st \ba {\cal O}} \tmn(f)\dsm
  -\int_{\ptz \ba {\cal O}}\tmn(\fb)\dsm   \\
  & + & \int_{\cal O} \left[ \tmn(f) - \tmn(\fb) \right] \dsm
  \ + \ \pu.
\end{array}\right\}\label{Born-Fermi}
\end{equation}

The Born field $f^B$ above is computed assuming that the proper
time $\tau$, $\ub$, $\ab$ and ${\bf e}_{(l)}$ at $A$ for both
particles (real and uniformly accelerated) coincide. Thus they
have the same hyperplane $\st$ passing through $A$ and the same
Fermi coordinates $x^l$ on it.

The right-hand side of \rf{Born-Fermi} does not depend, obviously,
upon a choice of ${\cal O} \subset \ptz$. On the grounds of
symmetry we must have: $\pu=m(a)u^{\nu}+p(a)a^{\nu}$, where $m(a)$
and $p(a)$ are phenomenological functions of one variable
$a=|\ab|$. We call (\ref{Born-Fermi}) the {\em Born-renormalized}
four-momentum of the system ``point particle + Maxwell field''.

Unfortunately, the above definition cannot be directly generalized
to the case of many particle system because, in general, there is
no common rest-frame space $\st$ for different particles. In what
follows we shall rewrite the above definition in a way, which
admits an obvious generalization to the case of many particles.
For this purpose we replace $\st$ by an arbitrary spacelike
hypersurface $\Sigma$ which is flat at infinity. More precisely,
one has

{\em Proposition.} Quantity (\ref{Born-Fermi}) may be rewritten as
follows:
 \be\left.\begin{array}{l}
\pbn=\int_{\Si\ba\oo}\tmn(f)\dsm+\int_{\oo}
[\tmn(f)-\tmn(\fb)]\dsm\\ -\int_{P\ba\oo} \tmn(\fb)\dsm + \pu
\end{array}\right\} \et{11} \ee
where $\Sigma$, $P$ are any space--like hypersurfaces which
coincide along some region $\oo$ around $A$ (we assume that $P$
has boundary equal to $\mathfrak{p}=\{z^3=-\frac1a\}$, $P$
approximates $\ptz$ at infinity and that $\Sigma$ approximates a
space-like hyperplane at infinity)-cf. the figure below.

$$ 
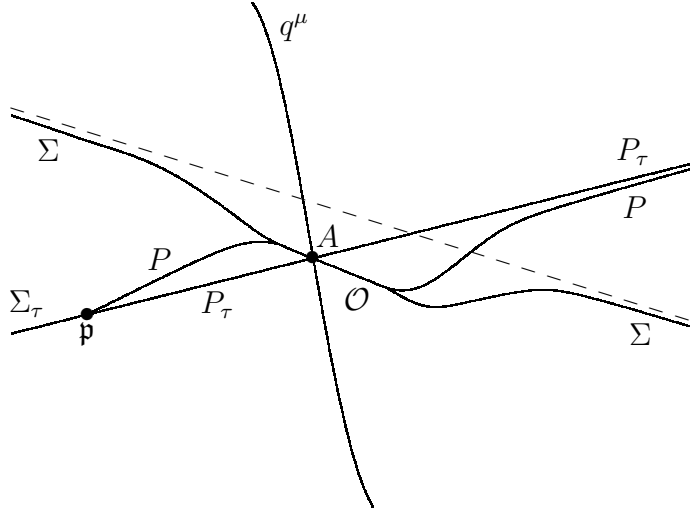
\xy {(-40,-10); (50,12.5) **\crv{ (-30,-7.5)*{\bullet}&
(0,0)*{\bullet}} ?(.06)="a"  ?(.47)="b" ?(0.95)="c"} \POS
"a"+(0,3.5)*{\Sigma_{\tau}} \POS "b"+(0,-3)*{P_{\tau}} \POS
"c"+(0,3.5)*{P_{\tau}} \POS (2,3)*{A} \POS
(-30,-10)*{\mathfrak{p}}, {(-8,34); (8,-33) **\crv{ (-5,30) &
(0,0) & (5,-28) } ?(.15)="d"} \POS "d"+(3,3)*{q^{\mu}}, {(-40,19);
(50,-9) **\crv{(-35,17.4) & (-30,15.5) & (-20,12.7) & (-10,5.5) &
(-7.5,3.5) & (-5,2) & (0,0) & (10,-4) & & (15,-7) & (20,-6) &
(30,-3.7)  & (35,-4.7) & (40,-6)} ?(.03)="e" ?(.98)="f" ?(.23)="g"
?(.51)="h"} \POS "e"+(0,-3)*{\Sigma} \POS "f"+(0,-3)*{\Sigma}
\POS "h"+(0,-3)*{\cal O}, (-40,20); (50,-8.125) **@{--}, {(-5,2);
(-30,-7.5) **\crv{ (-7,2.4) & (-10,2) & (-15,0)} ?(.4)="j"} \POS
"j"+(-11,-2)*{P}, {(10,-4); (50,11.8) **\crv{ (12,-4.5) & (15,-4)
& (20,0) & (25,4) & (30,6) & (35,7.5) & (40,9) } ?(0.95)="k"} \POS
"k"+(0,-3)*{P},
\endxy
$$

{\em Definition.} Hypersurface $\Sigma$ as in the Proposition is
called special if $\Sigma$ coincides with $\st$ in a neighbourhood
of $A$,  i.~e.~if one can take $P=\ptz$ (cf. the Noether theorem
$\partial_{\mu}T^{\mu\nu}=0$).

{\em Idea of proof.} First let $\Sigma$ be special ($P=\ptz$) and
choose $\oo$ contained in $\ptz\cap\Sigma$. Then the first terms
in \rf{Born-Fermi} and \rf{11} coincide ($f$ is a solution of
Maxwell equations, we use Noether theorem) and \rf{11} holds. We
can assume that in Fermi coordinates $\oo=K(A,R)$, a ball with a
small radius $R$. Next we can take  any  $\te\Si$, $\te P$,
$\te\oo$ as in the Proposition and denote the corresponding right
hand side of \rf{11} by $\te\pbn$. We need to prove
$\pbn=\te\pbn$. Now  we modify the interior of $\oo$, thus
replacing $\oo$ by $\te\oo^{\prime}$ without changing its
boundary, in such a way that small pieces of $\te\oo^{\prime}$ and
$\te\oo$ around $A$ coincide. It modifies $\pbn$ by
\[ \left|\left(\int_{\oo}-\int_{\te\oo^{\prime}}\right)
[\tmn(f)-\tmn(\fb)] \dsm\right| \leq 2\int_0^R Cr^{-2}r^2dr=2CR \]
(cf. Appendix B, $C$=const.). Next we replace $\te\oo^{\prime}$ by
its small piece contained in $\te\oo$. Finally, we modify $\Si$
and $P$ outside of that small piece getting $\te\Si$ and $\te P$,
which doesn't change $\te\pbn$ because $f,\fb$ are solutions of
the Maxwell equations (cf. the Noether theorem and the assumption 
before \rf1).
 Thus $\pbn$
for $\Si$ and $\te\pbn$ for $\te\Si$ differ by a term of order
$R$. Taking the limit $R\ra0$, we get $\te\pbn=\pbn$.

\section{Explicit formula for the four--momentum}

Here we specify  the hypersurface $\Si$ in \rf{11} to be  special
(i.e. $A\in\oo\su\Si\cap\st$), $P=\ptz$ (cf. Section 4) and choose
the spherical coordinates related to a Fermi--propagated system as
in Section 3.

Let $\uc\su\oo$ be given by $r<\rtf$. According to \rf{4a},
\rf{5c} and \rf9,
\[ \int_{\ptz\ba\uc}\tmn(\fb)\dsm  =\egd\nu0
 \frac{e^2}{2\pi^2}\is \{\frac1{16\rtf}\]
\[ +\frac18
a\cos\theta\lar+O(R)\} \sin\theta d\theta d\phi. \] Using
\rf{Born-Fermi},\rf{4a} and \rf{10a}, the Born--renormalized
four--momentum \be \pbn=\int_{\Si\ba\oo}\tmn(f)\dsm+
[\egd\nu{\lam} K^{(\lam)}(\oo)+\pu], \et{14}\ee where
\be\left.\begin{array}{l}
K^{(0)}(\oo)=\lim_{\uc\ra0}[\int_{\oo\ba\uc}\dbp d\Si \\
-\frac{e^2}{2\pi^2}\is \{\frac1{16\rtf} +\frac18 a\cos\theta\lar\}
\sin\theta d\theta d\phi] \\
=\int_{\oo}\{\dbp-\dss\}\ds-\int_{\st\ba\oo}\dss
\ds,\end{array}\right\}
 \et{15}\ee
\be K^{(k)}(\oo)=\int_{\oo}\dbk\ds. \et{16}\ee Now $\oo$ doesn't
need to be inside $\ptz$ -- only inside $\st$ -- use \rf{4a} and
\rf{15}-\rf{16}.

\section{Relation with the Coulomb--renormalization}
\label{relation}

According to \rf2 and \rf{11}, the difference between Born-- and
Coulomb--renormalized four--momentum
\[ \pbn-\pcn=-mu^{\nu}+L^{\nu}, \]
where
\[ L^{\nu}=\pu+\int_{\Si\ba\oo}\tmn(\fc)\dsm \]
\[ +P\int_{\oo}[\tmn(\fc)-\tmn(\fb)]\dsm-
\int_{P\ba\oo}\tmn(\fb)\dsm,\] which looks like \rf{11} but with
the $P$ sign. Repeating the arguments of Section 5, we get for
$L^{\nu}$ an analogue of \rf{14}--\rf{16}, again with the $P$ sign
and with $D$, $B$ replaced by $D^C$, $B^C$. Setting $\oo=\st=\Si$
and using \rf{1b},\rf{10p}, one obtains
\[ L^{\nu}-\pu=\lim_{R\ra0}\int_{\st\ba
K(R)}\{\frac12(D^C)^2-\dss\}\ds=0.\] Thus one gets

{\em Proposition.} Coulomb-- and Born--renormalization of
four--momentum give always the same result iff $\pu\equiv
mu^{\nu}$.

\section{Born--renormalization of the angular momentum tensor}

In analogy with \rf{Born-Fermi} we define the Born--renormalized
tensor of angular momentum \be \left.\begin{array}{l}\mbnl:=
\int_{\Si\ba\oo}\mmnl(f)\dsm+\int_{\oo}[\mmnl(f)-\mmnl(f^B)]\dsm
\\
-\int_{P\ba\oo}\mmnl(f^B)\dsm+\mua +\frac{e^2}{8\pi
a}(u^{\nu}a^{\lambda}-u^{\lambda}a^{\nu}),
\end{array}\right\}\et{17}\label{AM}\ee
where ${\cal M}$ was defined in \rf4, $\Sigma=\Sigma_{\tau}$,
$P=P_{\tau}$.

The above formula renormalizes the field infinity near the
particle, leaving opened the standard convergence problems at
spatial infinity ($r \rightarrow \infty$). We discuss briefly
these issues in Appendix A. Here, we only mention that these
global problems never arise, when the particle's equations of
motion are derived from the momentum and the angular momentum
conservation. Indeed, the conservation condition may always be
verified {\em locally}, i.~e.~on a family of hypersurfaces
$\te\Sigma_{\tau}$ which coincide outside of a certain (spatially compact)
world tube $T$. Comparing the value of angular momentum calculated
on two different $\te\Sigma_{\tau}$ never requires integration outside of
$T$, because the far-away contributions are the same in both
cases.

The last term in \rf{17} could be
incorporated into $\mua$ but for the future convenience (see
remark at the end of this Section) it was written separately.
The sum of those two terms can be interpreted as the total
angular--momentum of the particle dressed with the Born field. On
the symmetry grounds $ {\bf
M}(\ub,\ab)=(\ub\we\ab)R(a)+(\ub\we\ab)^*S(a) $ (one has
$(\ub\we\ab)^{\nu\lam}=u^{\nu}a^{\lam}-u^{\lam}a^{\nu}$, $
(\ub\we\ab)^*=(\ebd0\we a{\bf b}_{(3)})^*=a{\bf b}_{(1)}\we{\bf
b}_{(2)}$, cf. Section 3). 
Clearly \rf{17} doesn't depend on the choice of $\oo\subset P$.
Using the Appendices and the Noether theorem $\partial_{\mu}{\cal
M}^{\mu\nu\lambda}=0$, one proves \rf{17} for general $\Sigma$,
$P$ as in Proposition of Section 4.

If we restrict ourselves to special hypersurfaces, then using
\rf{4b}, \rf{10} and relation between $x^k$ and $z^k$ on $\st$
(Section 3), we get
\[ \int_{P\ba{\cal U}}\te\mmnl(f^B)\dsm=(\egd\nu l\egd\lam
0-\egd\lam l\egd\nu 0) \]
\[ \times
\left[\frac{e^2}{32\pi^2}  \is \{-\lar \frac{x^l}r +O(R)\}
\sin\theta d\theta
 d\phi
-\frac{e^2a^l}{8\pi a}\right], \] where ${\cal U}$ is given by
$r<\rtf$. Next,  \rf{17}, \rf{4b} and \rf{10b} give (uncontinuous
terms of $\frac{a^{\lambda}}{a}$ type cancel out!)
\be\left.\begin{array}{l} \mbnl=\int_{\Si\ba\oo} \mmnl(f)\dsm+
(\egd\nu l\egd\lam\rho-\egd\lam l\egd\nu\rho) L^{(l)(\rho)}(\oo)\\
+[q^{\nu}(\tau)\egd\lam\rho-q^{\lam}(\tau)\egd\nu\rho]K^{(\rho)}(\oo)+
\mua,
\end{array}\right\}\et{18} \ee
where \be\left.\begin{array}{l} L^{(l)(0)}(\oo)= \lim_{{\cal U}\ra
0}[ \int_{\oo\ba{\cal U}} \frac12 x^l(D^2+B^2)\ds+ \\
\frac{e^2}{32\pi^2}  \is \lar \frac{x^l}r  \sin\theta d\theta
 d\phi]\\ = \int_{\oo}\left\{\frac12 x^l
(D^2+B^2)-\left[\frac12 x^l(D^s)^2\right]_s\right\}\ds\\
-\int_{\Sigma_{\tau}\ba\oo} \left[\frac12 x^l(D^s)^2\right]_s\ds,
\end{array}\right\}
 \et{19} \ee
\be L^{(l)(k)}(\oo)=\int_{\oo}x^l\dbk\ds. \et{20}\ee Again $\oo$
doesn't need to be inside $P$.

Finally, comparing on $\Si_{\tau}$ the Born-- and
Coulomb--renormalization, i.e.
\[ \mcnl=P\int_{\Si_{\tau}}[\mmnl(f)-\mmnl(f^C)]\dsm \]
(cf. Appendix A), we get that $\mbnl-\mcnl$ equals \rf{17} for
$f=f^C$ and the $P$ sign before the second term. Thus it equals
\rf{18} with $D,B$ in \rf{15}, \rf{16}, \rf{19}, \rf{20} replaced
by $D^C$, $B^C$ and the $P$ sign in \rf{15}. Setting
$\oo=\Si_{\tau}=\Si$ and using \rf{1b} and \rf{10aa},
$\mbnl-\mcnl=\mua$. Therefore both renormalizations give the same
result iff $\mua=0$. This equation was the reason to separate the
last term in Definition (\ref{AM}) from $\mua$.

\section{The case of many particles}

Here we extend the results of Sections 4--7 to the case of many
particles. For the $i$-th particle we fix a point $A_i$ on its
trajectory, corresponding to a proper time $\tau_i$. The
space--like hypersurface $\Si$ passes through all $A_i$. At the
beginning we assume that  some regions $\oi\su\Si$ around $A_i$
are  contained in $\pti\su\sti$ corresponding to the $i$-th
particle and that, asymptotically, $\Sigma$ approximates a
space-like hyperplane (special hypersurface). The formula for the
Born--renormalized four--momentum generalizes to:
\be\left.\begin{array}{l}
\pbn=\int_{\Si\ba\cup\oi}\tmn(f)\dsm+\sum_i\int_{\oi}
[\tmn(f)-\tmn(\fbi)]\dsm\\ -\sum_i\int_{\pni\ba\oi} \tmn(\fbi)\dsm
+ \sum_i\pua, \end{array}\right\} \et{22} \ee $\pni=\pti$ for the
$i$-th particle. One proves the analogue of the Proposition of
Section 4. For the special hypersurface \rf{14} generalizes to:
\be \pbn=\int_{\Si\ba\cup\oi}\tmn(f)\dsm+ \sum_i[\egd\nu{\lam}
K^{(\lam)}(\oi)+\pua], \et{24}\ee where $K^{(\lam)}(\oi)$ are
given as in \rf{15}--\rf{16} with $\ui$ described by $r<\ritf$.

The comparison with the Coulomb renormalization gives
\[ \pbn-\pcn=\sum_i\{-m_{(i)}u^{\nu}_{(i)}+L^{\nu}_{(i)}\}, \]
where $L^{\nu}_{(i)}$ are as in Section 6 and the analogue of the
Proposition of Section 6 holds.

The Born--renormalized tensor of angular momentum takes now the
form: \be \left.\begin{array}{l}\mbnl=
\int_{\Si\ba\cup\oi}\mmnl(f)\dsm+\sum_i\int_{\oi}[\mmnl(f)-\mmnl(\fbi)]\dsm
\\ -\sum_i\int_{P_{(i)}\ba\oi}\mmnl(\fbi)\dsm+ \sum_i
\left[\md\nu\lam_{(i)}(\ub_{(i)},\ab_{(i)}) +\frac{e_{(i)}^2}{8\pi
a_{(i)}}(\ub_{(i)}\we\ab_{(i)})^{\nu\lambda}\right].
\end{array}\right\}\et{25}\ee
The analogue of Proposition of Section 4 holds.
What concerns \rf{18}, $\Si\ba\oo$ is now replaced by
$\Si\ba\cup\oi$ and one has $\sum_i$ before the remaining
$i$--dependent terms. For a special $\Si$ one defines the Coulomb
renormalization \be
\mcnl=P\int_{\Si}[\mmnl(f)-\sum_i\mmnl(\fci)]\dsm \et{26}\ee and
shows that the both renormalizations give the same result iff
$\md\nu\lam_{(i)}(\ub_{(i)},\ab_{(i)})\equiv 0$ for all $i$.

\begin{appendix}

\section*{Appendices}
\renewcommand{\theequation}{\Alph{section}.\arabic{equation}}

\section{Field fall-off conditions at spatial infinity and a
possibility to define global angular momentum} \label{fall-off}

To define the four-momentum of the system, we assume that 
the field behaves at spatial infinity (i.~e.~for $r\ra\infty$) 
like $r^{-2}$. To define globally the angular momentum 
of the system, much stronger fall-off conditions
are necessary. Here, we present a possible choice: 
we assume
that the field behaves at spatial infinity like a superposition 
of boosted Coulomb
fields (modulo $r^{-3}$--terms). Then (for any space-like 
hyperplane)  the
angular momentum density behaves like an anti-symmetric
$r^{-3}$--term (modulo $r^{-4}$--terms). This is
sufficient to define global value of angular momentum using the
``principal value'' sign for integration at infinity. This means
that we first integrate over spatially symmetric regions $V
\subset \Sigma$ of an asymptotically flat hypersurface $\Sigma$
and then pass to the limit $V \rightarrow \Sigma$. The symmetry
depends upon a choice of a central point $x_0$, but it is easy to
check that the final result of such a procedure {\em does not}
depend upon this choice. Moreover, the above asymptotic conditions
allow us to change $\Sigma$ at infinity. Indeed, the difference
between results obtained for different $\Sigma$'s equals to
a surface  integral at infinity which vanishes as a consequence of
the assumed asymptotic conditions (cf. also \cite{Al}). We stress,
however, that the renormalization proposed in the present paper
cures the local and not global problems. Derivation of particle's
equations of motion from field equations does not rely on the
global problems.


\section{Approximation by the Born field near the trajectory}

Suppose (cf. Section 4) that we have two trajectories: of a real
particle $p$ and of the reference particle $\te p$, which is
uniformly accelerated. Both trajectories touch at $A$, where the
proper times $\tau_0$, the four--velocities $\ub$, the
accelerations $\ab$ and ${\bf e}_{(l)}$ coincide. In general, the
quantities related to $\tilde p$ differ from those related to $p$
and are distinguished by tilde. Then
$A\in\sd{\tau_0}=\te\sd{\tau_0}$, but for $H$ approaching $A$, one
has $H\in\sd\tau\cap\te\sd{\te\tau}$ and in general
$\tau\neq\te\tau$. Denote by $r$ ($\te r$) the radius of $H$
w.r.t. $\sd\tau$ ($\te\sd{\te\tau}$). One has

{\em Proposition.} Suppose that $H$ belongs to a region of
space--like directions w.r.t. $A$, which is separated from the
light cone at $A$ and that $H$ approaches $A$. Then
\begin{enumerate}
\item[1)] $r/\te r,\te r/r\sim 1$
\item[2)] $\tmn(f)-\tmn(f^B)\sim r^{-2}$
\item[3)] $\mmnl(f)-\mmnl(f^B)\sim r^{-2}$
\end{enumerate}
where $f$ is a Maxwell field related to $p$, $f^B$ is the Born
solution related to $\te p$.

{\em Idea of proof.} We may set $\tau_0=0$, $q(0)=0$,
$u(0)=(1,0,0,0)$, $a(0)=\te a(0)$. One has $\dd a(\tau)\equiv
a(\tau)-\te a(\tau)\sim\tau$, $\dd u(\tau)\sim\tau^2$, $\dd
q(\tau)\sim\tau^3$, the angle between $\sd\tau$ and
$\te\sd{\te\tau}$ is of order $\tau^2$. Denoting by $\de F$ the
difference of $F$ computed for $H$ w.r.t. $\sd\tau$ and w.r.t.
$\te\sd{\te\tau}$ and using geometric considerations, we  get
$\tau\leq Cr$, $\de\tau\sim r^3$, $\de x^k\sim r^3$, $\de r\sim
r^3$, $\te r/r-1=\de r/r\sim r^2$, 1) follows. Using
$\tmn=\egd\mu\al\egd\nu\beta\tgn\al\beta$, \rf3 and \rf1, one gets
$ \tmn\sim\egd\mu\al\egd\nu\beta(r^{-2}+ar^{-1}+C)^2\sim r^{-4}$,
$\de\dt{\egd\mu\al}\sim\de a\sim\tau$,
$\de{\egd\mu\al}\sim\tau^2\sim r^2$, $\de r^{-2}\sim r^{-3}\de
r\sim 1$, $\de a\sim r$, $\de r^{-1}\sim r^{-2}\de r\sim r$, $\de
T\sim r^{-2}$, 2) holds. Moreover, $\de q^{\lam}=\Delta
q^{\lam}+\te q^{\lam}(\tau)- \te
q^{\lam}(\te\tau)\sim\tau^3+r^3\sim r^3$,
\[ \de y^{\lam}=\de q^{\lam}+x^l\de\egd\lam l+(\de x^l)\egd\lam l\sim
r^3+r\cdot r^2+r^3\cdot 1\sim r^3, \]
\[ \de\mmnl= y^{\nu}\de T^{\mu\lam}+(\de y^{\nu})T^{\mu\lam}
-(\lam\leftrightarrow\mu)\sim 1\cdot r^{-2}+r^3\cdot r^{-4}\sim
r^{-2}, \]
 3) follows.

\end{appendix}

\begin{center} {\bf Acknowledgements} \end{center}

We thank Marcin Ko\'scielecki and Szymon Charzy\'nski for fruitful
discussions.

\end{document}